\documentclass[aps,superscriptaddress]{revtex4}
\usepackage{amssymb,amsmath,epsfig}
\usepackage[colorlinks=true, pdfstartview=FitV, linkcolor=blue, citecolor=red, urlcolor=magenta, breaklinks=true]{hyperref}
\usepackage{graphicx}
\usepackage{epstopdf}

\begin{document}
\title{The (de)-confinement transition in tachyonic matter at finite temperature}

\author{Adamu Issifu}\email{bigadamsnet@yahoo.com}
\affiliation{Departamento de F\'{\i}sica, Universidade Federal de Campina Grande
Caixa Postal 10071, 58429-900 Campina Grande, Para\'{\i}ba, Brazil}

\author{Francisco A. Brito}\email{fabrito@df.ufcg.edu.br}
\affiliation{Departamento de F\'{\i}sica, Universidade Federal de Campina Grande
Caixa Postal 10071, 58429-900 Campina Grande, Para\'{\i}ba, Brazil}
\affiliation{Departamento de F\'isica, Universidade Federal da Para\'iba, 
Caixa Postal 5008, 58051-970 Jo\~ao Pessoa, Para\'iba, Brazil}

\begin{abstract} 
{In this paper we present a QCD motivated model that mimics QCD theory.} 
We examine the characteristics of the gauge field coupled with the color dielectric function ($G$) in the presence of temperature ($T$). The aim is to achieve confinement at low temperatures $T<T_{c}$, ($T_{c}$, is the critical temperature), similar to what occurs among quarks and gluons in hadrons at low energies. {Also, we investigate scalar glueballs and QCD string tension and effect of temperature on them.} To achieve this, we use the phenomenon of color dielectric function in gauge fields in a slowly varying tachyon medium. This method is suitable for analytically computing the resulting potential, glueball masses and the string tension associated with the confinement at a finite temperature. We demonstrate that the color dielectric function changes Maxwell's equation as a function of the tachyon fields and induces the electric field in a way that brings about confinement during the tachyon condensation below the critical temperature. 
\end{abstract}

\maketitle
\pretolerance10000

\section{Introduction}
Quantum chromodynamics (QCD) is a theory that attempts to explain the strong interactions carried by gluons that keep quarks and gluons in a confined state in hadrons. The success of this theory depends on asymptotic freedom \cite{1,2,3}. QCD forms the bases of nuclear physics and enable us to appreciate and explain the features of matter. 
{However, non-relativistic perturbative QCD theories cannot accurately reproduce the results of charmonium and bottomonium spectra outputs, unless the leading {\sl renormalon} terms cancels out. In this case, the net energy of such bound states from QCD potentials, are in agreement with phenomenological potentials for the range 0.5 GeV$^{-1}$ $\lesssim r \lesssim$ 3 GeV$^{-1}$ \cite{Sumino:2001eh}. Thus, for many years now, it has been conceived that phenomenological potential models are described by such systems.

It has been realized that owing to the gluon confinement, the QCD vacuum shows a characteristic of a dielectric medium \cite{QCD vacuum:1}. This idea has been employed in developing several models, including MIT bag model \cite{MIT bag}, SLAC bag model \cite{SLAC bag}, Cornell potential for heavy quarks \cite{Conell potential} and many soliton models \cite{others,Phenomenology,guendelman} which are used to describe hadron spectroscopy. However, until recently, no successful effort had been made to compute the color dielectric function representing the QCD vacuum from quantum theory. In view of this, all the models developed using the dielectric function approach were considered phenomenological  
though it agrees with QCD as shown in \cite{Reinhardt:2008ek}.}

There exists some similarity between QCD and QED (\textit{Quantum Electrodynamics}) in terms of the successes of both theories, 
but they depart from each other by their strength, medium and dynamics of interactions. QED explains the interaction between charged particles while QCD explains the strong interaction between sub-atomic particles. QED creates a \textit{screening effect} which decreases its net electric charge as inter-(anti-)particle distance increases. The opposite effect is observed in QCD where a \textit{anti-screening} occurs, the net color-charges increase with increasing distance between (anti-)particle pairs. This similarity and the differences make it interesting to advance a study of one in terms of the other. In this work we will explore QCD in terms of QED. The most popular potential for heavy quarks at confined state is the Cornell potential known to be, $v_{c}=-\dfrac{a}{r}+br$, where $a$ and $b$ are positive constants. This potential comprises linearly increasing part (infrared interaction) and Colombian part (ultraviolet interaction) \cite{2,7}.  

In this paper, we will establish that the phenomenon of confinement is achievable with an electric field immersed in a color dielectric medium ($G$) at a finite temperature $T$. We will also show that the net potential resulting from the confinement at $T=0$ is similar to Cornell's potential for confinement of heavy quarks and gluons in hadrons at low energies.  {Our dielectric function $G$ is identified with tachyon condensation \cite{6} at low energies.} The tachyon matter creates the necessary conditions for confinement phase at low temperatures ($T<T_{c}$) and de-confinement phase at high temperatures ($T\geq T_{c}$). The relevance  of the color dielectric function $G(\phi)$ is to generate the strong interaction needed for (de-)confinement of the associated colored particles \cite{4,5,6}. 

Many works have been done on determining the potentials for quark confinement as a function of temperature, commonly called thermal QCD, by using a number of different approaches including Wilson and Polyakov loop corrections %
\cite{8,9,10,11}. 
Most of the challenges posed by these models stems from the proper behavior of the QCD string tension at all temperatures as compared with lattice simulation results. The expected behavior of the string as suggested by many simulation results is; a sharp decrease with temperature at $T<T_{c}$, vanishes at $T=T_{c}$ and slowly decreases at $T>T_{c}$ \cite{1,2,11,13}. 

The main purpose of this paper is to determine analytically, the net static potential for the {quarks and gluons confinement} in $3+1$ dimensions as a function of temperature. {We will also obtain the QCD string tension and glueball masses associated with it as a function of temperature and study its behavior.} 
We will use an Abelian theory, as it is applied to QED, but the dielectric function $G(r)$ will be carefully chosen to give the expected (de-)confinement in the chosen tachyon matter \cite{14,15}. It has already been shown that the Abelian part of the non-Abelian QCD string tension constitutes $92\%$, that comprises linear part of the net potential. Hence, we can estimate non-Abelian theory using an Abelian approach \cite{16,17}. This fact also permits us to study the QCD theory phenomenologically to establish the confinement of the quarks and gluons inside the hadron \cite{SLAC bag,18,MIT bag,21}.
The self-interacting scalar fields $\phi(r)$ describes the dynamics of the dielectric function in the \textit{tachyon matter}.  
Thus, we shall use a Lagrangian that would collectively carry information on the dynamics of the gauge, the scalar field associated with the tachyon dynamics and temperature.

{The motivation for using this approach is twofold. Firstly,  because we are able to  study QCD phenomenologically by identifying the color dielectric function naturally with the tachyon potential. Secondly, one can apply such phenomenological approach to obtain models that mimic QCD in stringy models where temperature effects in tachyon potentials  \cite{Hotta:2002nt} can be considered in brane confinement scenarios \cite{brane}. In this case, it may also bring new insight to confining supersymmetric gauge theories such as the Seiberg-Witten theory \cite{dual} that deals with electric-magnetic duality and develops magnetic monopole condensation.

Thus, we choose a tachyon potential which is expected to condense at some value \cite{26} at the same time that the gauge field is confined.
This phenomenon coincides with the dual Higgs mechanism, where the dual gauge field becomes massive \cite{brane}. This means that in the infrared the QCD
vacuum is a perfect color dielectric medium and therefore a dual superconductor in which magnetic monopole condensation leads to electric field confinement \cite{dual}.

The paper is organized as follows. 
{In Sec.~\ref{sec2} and ~\ref{rgd} we review both the theory of electromagnetism in a dynamical dielectric medium and gluodynamics, with its associated QCD-like vacua respectively. In Sec.~\ref{sec3} we introduce the tachyon Lagrangian coupled with temperature and its associated effective potential. In the same section, we study glueball masses at zero temperature ($T=0$) and at a finite temperature ($T$) and analyze their characteristics.} In the latter cases we find analytically the net potential for {confinement of quarks and gluons} as a function of temperature. We analyze   the characteristics of the net confinement potential. Also, we analyze the QCD string tension as a function of temperature. In Sec.~\ref{sec4} we present our final comments.   

\section{Maxwell's equations modified by dielectric function}
\label{sec2}

In this section we will review the theory of electromagnetism in a {\textit{color dielectric medium}} to set the pace for us to explain the phenomenon of confinement. Beginning with the  Maxwell Lagrangian with no sources we have
\begin{equation}\label{1}
\mathcal{L}=-\dfrac{1}{4}F_{\mu\nu} F^{\mu\nu}.
\end{equation}
Its equations of motion are
\begin{equation}\label{2}
\partial_\mu F^{\mu\nu}=0.
\end{equation}
It is worth mentioning that though the Lagrangian is with no source, its equations of motion still admit solutions with spherical symmetry \cite{24}.

Consider the gauge field in dielectric medium, $G(\phi)$, with $\phi$ the field describing the dynamics of the medium. The Lagrangian above can be rewritten as
\begin{equation}\label{3}
\mathcal{L}=-\dfrac{1}{4} G(\phi) F_{\mu\nu} F^{\mu\nu}.
\end{equation}
Its equations of motion are 
\begin{equation}\label{4}
\partial_{\mu}[G(\phi) F^{\mu\nu}]=0.
\end{equation}
Let us impose the restrictions, $\mu=1,2,3$ and $\nu=0$. Thus,
\begin{equation}\label{5}
\nabla.[G(\phi)\textbf{E}]=0.
\end{equation} 
The magnetic field is not of interest in this work, because our concentration will be on the electric field confinement only, so the indices were deliberately defined to avoid the magnetic field. 

We begin with Eq.~(\ref{5}) to determine the solution of the electric field $\textbf{E}$ in the dielectric medium $ G(\phi)$. As stated above, all the solutions will be computed in spherical symmetry, i.e. $E(r)$ and $\phi(r)$ are only radially, $r$, dependent, hence $G(\phi)$ follows the same definition, thus
\begin{equation}\label{6}
\nabla.[G(\phi)\textbf{E}]=\dfrac{1}{r^{2}} \dfrac{\partial}{\partial r} (r^{2} G(\phi) E_{r})=0,
\end{equation}
and
 \begin{equation}\label{7}
 E_{r} =\dfrac{\lambda}{r^{2} G(\phi)}.
 \end{equation}
 Here, $\lambda$ is the integration constant which can be related to electric charge as $\lambda=\dfrac{q}{4\pi\varepsilon_{o}}$. Therefore, the electric field solution, $E$ in the dielectric medium $G(\phi)$ can be represented as
 
 \begin{equation}\label{8}
 E= \dfrac{q}{4\pi \varepsilon_o r^{2} G(\phi)},
 \end{equation}
where $E=\mid\textbf{E}\mid=E_r$. Consequently, the dielectric medium changes the strength of the $\textbf{E}$ as a function of $\phi$. 

The coupling between electromagnetism and scalar field dynamics at finite temperature is given by the effective Lagrangian
 \begin{equation}\label{9}
\mathcal{L}= - \dfrac{1}{4} G(\phi) F_{\mu\nu} F^{\mu \nu} +\dfrac{1}{2} \partial_{\mu} \phi \partial^{\mu} \phi -V_{eff}(\phi).
\end{equation}
Its effective potential as a function of the scalar field, $\phi$, at a finite temperature, $T$, has already been found and is given by \cite{jackiw,weinberg}
\begin{equation}\label{10}
V_{eff}(\phi)=V(\phi)+\dfrac{T^{2}}{24} V_{\phi \phi}(\phi),
\end{equation}
where $V_{\phi \phi}$ is the second derivative of $V(\phi)$.

The behavior of the dielectric function $G(\phi)$ will be obtained from the equations of motion \cite{25} of the above Lagrangian. The equations of motion for the various fields, i.e., the gauge field ${A_{\mu}}$ and the scalar field $\phi $, found in the above Lagrangian are given as
\begin{equation}\label{11}
\partial_{\mu}[G(\phi)F^{\mu\nu}]=0,
\end{equation}
and
\begin{equation}\label{12}
\partial_{\mu} \partial^{\mu} \phi + \dfrac{1}{4}\dfrac{\partial G(\phi)}{\partial\phi} F_{\mu\nu} F^{\mu\nu}+\dfrac{T^{2}}{24}\dfrac{\partial V_{\phi\phi}}{\partial\phi}+\dfrac{\partial V(\phi)}{\partial\phi}=0.
\end{equation}  
The equations of motion for the scalar field $\phi$ and the gauge field $A_\nu$ with radial symmetry are
\begin{equation}\label{13}
\dfrac{1}{r^{2}}\dfrac{d}{d r} (r^{2} G(\phi)E)=0,
\end{equation}
and
\begin{equation}\label{14}
\dfrac{1}{r^{2}}\dfrac{d}{d r} \left( r^{2}\dfrac{d \phi}{d  r}\right) =\dfrac{T^{2}}{24} \dfrac{\partial V_{\phi\phi}}{\partial\phi}-\dfrac{1}{2}\dfrac{\partial G(\phi)}{\partial\phi} E^{2} +\dfrac{\partial V(\phi)}{\partial\phi}=0.
\end{equation}
As has been shown above we can identify  that the solution of  Eq.~(\ref{13}) is that given by Eq.~(\ref{8}).

To establish strong interaction and its resultant confinement, our dielectric function 
needs to asymptotically satisfy these  conditions:
\begin{equation}\label{15}
G(\phi(r))= 0
\qquad\text{as}\qquad
r\rightarrow r_*,
\end{equation}
and
\begin{equation}\label{16}
G(\phi(r))= 1
\qquad\text{as}\qquad
r\rightarrow 0,
\end{equation}
where $r_*$ stands for the scale where the confinement starts to become effective. Particularly, $r_*=\infty$ for $G(\phi(\infty))\sim \frac{1}{r^{2}}$ and from Eq.~(\ref{8}) we find $E\equiv\textit{constant}$. This uniform electric field behavior agrees with confinement  \textit{everywhere}.

{\section{gluodynamics and QCD-Like vacuum}\label{rgd}

In this section we analyze the gluodynamics in the tachyon matter. The Lagrangian for gluodynamics is given as 
\begin{equation}\label{gd1}
\mathcal{L'}=-\frac{1}{4}F^a_{\mu\nu}F^{a\mu\nu}+|\epsilon_v|,
\end{equation}
where $-|\epsilon_v|$ represents the vacuum energy density that keeps the scale and the conformal symmetries of gluodynamics broken. 
Gluodynamics is generally known to be scale and conformally invariant in the limit of classical regime but the symmetry breaks down when there is quantum effect due to non vanishing gluon condensate $\langle F^a_{\mu\nu}F^{a\mu\nu}\rangle > 0$ \cite{gluodynamics}. This is what brings about the anomaly in the QCD energy-momentum tensor ($\theta^{\mu\nu}$) trace
{\begin{equation}\label{gb1a}
\theta^\mu_\mu=\frac{\beta(g)}{2g} F^a_{\mu\nu}F^{a\mu\nu}.
\end{equation}}
The leading term of the $\beta$-function of the coupling $g$ is given by
{\begin{equation}\label{gd1b}
 \beta(g)=-\frac{11g^3}{(4\pi)^2},
 \end{equation}}
with the vacuum expectation value given as 
\begin{equation}\label{gd2}
\langle\theta^\mu_\mu\rangle=-4|\epsilon_v|.
\end{equation}
The purpose of this section is to compute the energy-momentum tensor of Eq.~(\ref{9}) at $T=0$ and reconcile the results with Eq.~(\ref{gd2}) as applied in Ref. \cite{gluon-d}. This will require some cancellations of the tachyon fields and the gluon contributions to the vacuum density
\begin{equation}\label{gd3}
\theta^\mu_\mu=g^{\mu\nu}\left(2\dfrac{\partial \mathcal{L}}{\partial g^{\mu\nu}}-g_{\mu\nu}\mathcal{L}\right) +8\partial^2_\mu\phi.
\end{equation} 
The last term is the {total derivative} of the tachyon field, that is sometimes left out in the energy-momentum tensor computation. But it is sometimes necessary in quantum field theory due to some Ward identities \cite{gluodynamics 1}. Using the equation of motion Eq.~(\ref{14}) into Eq.~(\ref{gd3}) yields 
\begin{equation}\label{gd4}
\theta^\mu_\mu= G'(\phi)F^{\mu\nu}F_{\mu\nu}+4V'(\phi).
\end{equation}
Thus, we can relate Eq.~(\ref{gd2}) and Eq.~(\ref{gd4}) as
\begin{equation}\label{gd5}
\langle G'(\phi)F^{\mu\nu}F_{\mu\nu}\rangle=-4\langle|\epsilon_v|+V'(\phi)\rangle,
\end{equation}
where $G'(\phi)$ and $V'(\phi)$ represent the first derivative of the ``effective'' color dielectric function and the ``effective'' potential respectively. It is expected that in the classical limit $|\epsilon_v|\rightarrow 0$ the classical equation $\theta^\mu_\mu=0$ should be recovered. Therefore, we redefine the potential to include the vacuum energy density in the form \cite{gluon-d}
\begin{equation}\label{gd5a}
V(\phi)\to-|\epsilon_v|V(\phi).
\end{equation}
 Consequently Eq.~(\ref{gd5}) becomes
 \begin{equation}\label{gd6}
 \langle G'(\phi)F^{\mu\nu}F_{\mu\nu}\rangle=4|\epsilon_v|\langle V'(\phi)-1\rangle.
 \end{equation}
This equation guarantees the correct classical behavior where Eq.~(\ref{gd6}) vanishes at $|\epsilon_v|\rightarrow 0$ as expected. It is important to add that, with some quantum corrections we can obtain a non-vanishing contributions to $\langle\theta^\mu_\mu\rangle$ in the same limit. This result is similar to the results obtained in Ref.~\cite{gluon-d} for dilaton theory. We will show in the subsequent sections that the potential is precisely equivalent to the color dielectric function in string theory and thus, it represents the QCD vacuum density modified by a function $G(\phi)$. 
Using Eq.~(\ref{gd6}) and the Lagrangian in Eq.~(\ref{9}) we can redefine the effective potential to include the vacuum energy density as
\begin{equation}\label{gd6a}
V_{eff}\to-|\epsilon_v|V_{eff}(\phi,T)
\end{equation} 
Therefore, the gluon condensate for this potential becomes
\begin{equation}\label{gd7}
\langle G'(\phi)F^{\mu\nu}F_{\mu\nu}\rangle=4|\epsilon_v|\langle V_{eff}'(\phi,T)-1\rangle.
\end{equation} 
This equation also vanishes at $|\epsilon_v|\rightarrow 0$, {in consistency} with the classical prediction and we recover exactly Eq.~(\ref{gd6}) at $T=0$. We will soon find that the magnitude of $\langle G'(\phi)F^{\mu\nu}F_{\mu\nu}\rangle$ reduces at $T=T_c$ which is also expected. 
}

\section{Tachyon Condensation and Confinement}
\label{sec3}

In this section we will establish the relationship between tachyon condensation and confinement. Tachyons are particles that are faster than light, have negative masses and are unstable. Their existence are presumed theoretically in the same way as \textit{magnetic monopoles}. Tachyons just as magnetic monopoles have never been seen isolated in nature. In superstring theory, they are presumed to be interacting with other particles or interacting with each other at higher orders to form \textit{tachyon condensation} \cite{6,22}. Tachyon condensation is directly related to confinement just as monopole condensation.

\subsection{Tachyon Lagrangian with electromagnetic field and temperature}

From Eq.~(\ref{14}) we only have the  potential $V(\phi)$ and the dielectric function $G(\phi)$ as a functions of $\phi(r)$. Meanwhile, it will be convenient to restrict these choices as $G(\phi(r))=V(\phi(r))$. The propriety of this assertion will be demonstrated below in  awhile, working with a Lagrangian that characterizes the dynamics of the tachyon fields, $\phi(r)$.

To start with, let us consider the Lagrangian at Eq.~(\ref{9}) without the temperature correction as seen in \cite{5}
\begin{equation}\label{17}
\mathcal{L}= - \dfrac{1}{4} G(\phi) F_{\mu\nu} F^{\mu \nu} +\dfrac{1}{2} \partial_{\mu} \phi \partial^{\mu} \phi-V(\phi) .
\end{equation}
The equation of motion of this Lagrangian is given by
\begin{equation}\label{18}
\partial_{\mu} \partial^{\mu} \phi + \dfrac{1}{4}\dfrac{\partial G(\phi)}{\partial\phi} F_{\mu\nu} F^{\mu\nu}+\dfrac{\partial V(\phi)}{\partial\phi}=0.
\end{equation}
For simplicity let us consider the fields only in one dimension $x$, this yields,
\begin{equation}\label{19}
\phi=\phi(x)
\text{,}\qquad
A_{\mu} = A_{\mu}(x).
\end{equation}
The resulting equations of motion are 
\begin{equation}\label{20}
\dfrac{d}{d x} [G(\phi)E]=0,
\end{equation}

\begin{equation}\label{21}
-\dfrac{d^{2}\phi}{d x^{2}}-\dfrac{1}{2} \dfrac{\partial G}{\partial\phi} E^{2}+\dfrac{\partial V}{\partial\phi}=0,
\end{equation}
where we used  $F^{01}=E$. Integrating Eq.~(\ref{20}) we have 
\begin{equation}\label{22}
G(\phi) E = q 
\qquad{\Rightarrow}\qquad
E=\dfrac{q}{G(\phi)}.
\end{equation}
Substituting Eq.~(\ref{22}) into Eq.~(\ref{21}), we find
\begin{equation}\label{23}
-\dfrac{d^{2}\phi}{d x^{2}}-\dfrac{1}{2} \dfrac{\partial G}{\partial\phi} \dfrac{q^{2}}{G(\phi)^{2}}+\dfrac{\partial V}{\partial\phi}=0.
\end{equation}
We now make use of the tachyon Lagrangian commonly known in string theory with tachyon dynamics, $\tilde{T}(x)$, with electric field $E(x)$. Hence, for slowly varying tachyon fields, we can expand our Lagrangian in power series as  \cite{6,22}
\begin{align}\label{24-1}
e^{-1}\mathcal{L}&=-V(\tilde{T})\sqrt{1-\tilde{T}'^2 +F_{01}F^{01}}\\
&=-V(\tilde{T})[1-\dfrac{1}{2}(\tilde{T}'^2+F_{01} F^{01})+...]\nonumber \\
 &=-V(\tilde{T})+\dfrac{1}{2} V(\tilde{T})\tilde{T}'^2-\dfrac{1}{2} V(\tilde{T})F_{01}F^{01}+...\nonumber\\
 &=-V(\phi)+\dfrac{1}{2}\phi'^{2}-\dfrac{1}{2}V(\phi)F_{01}F^{01}+...,
 \label{24}
\end{align}
where $e=\sqrt{\mid{g}\mid}$ represents the general space-time. 
This relation also holds for $3+1$ dimensions for $\phi$ a function of $r$, which can be associated with $x$. In Eq.~(\ref{24}) we can relate   
\begin{align}\label{25}
V(\tilde{T}(\phi))= \left( \dfrac{\partial\phi}{\partial \tilde{T}}\right) ^{2}&\Longrightarrow\dfrac{1}{2}V(\tilde{T})(\tilde{T}'^2)
\nonumber\\
&=\dfrac{1}{2}\left( \dfrac{\partial\phi}{\partial\tilde{T}}\dfrac{\partial\tilde{T}}{\partial x}\right) ^{2}=\dfrac{1}{2}\phi'^{2},
\end{align} 
with $\phi=f(\tilde{T})$, or $\tilde{T}=f^{-1}(\phi)$. Now comparing Eq.~(\ref{24}) with Eq.~(\ref{17}) we find the equality $G=V$. This result is also true for Eq.~(\ref{9}) up to the thermal correction term. From the perspective of 
string theory, the thermal correction affects the tachyon potential $V(\tilde{T})$ of the original tachyon Lagrangian (\ref{24-1}) --- see e.g. \cite{Hotta:2002nt} and references therein. 
In our context we restrict ourselves to effective quantum field theory, where the one loop thermal corrections from the scalar sector affects $V(\phi)$ as given in the Lagrangian (\ref{9}).

\subsubsection{Confinement potential for the electric field in three dimensions as a function of temperature} 
For the tachyon Lagrangian in equation (\ref{24}), it is increasingly clear that the dielectric function $G(\phi)$ is equal to the potential $V(\phi)$. 
Now, we choose the appropriate classical tachyon potential that gives us the appropriate behavior for \textit{confinement and de-confinement} in the presence of temperature. We choose 

\begin{equation}\label{26}
V(\phi)=\frac{1}{2}[\alpha^{2}\phi^{2}-1]^{2}.
\end{equation}
In $3+1$ dimensions in radial coordinates, Eq.~(\ref{12}) can be rewritten as
\begin{equation}\label{27}
-\left[ \dfrac{1}{r^{2}}\dfrac{d}{d r} \left( r^{2}\dfrac{d \phi}{d  r}\right) \right] +\dfrac{T^{2}}{24} \dfrac{\partial V_{\phi\phi}}{\partial\phi}-\dfrac{1}{2}\dfrac{\partial G(\phi)}{\partial\phi} E^{2} +\dfrac{\partial V(\phi)}{\partial\phi}=0.
\end{equation}
Recall that the solution for the electric field is
\begin{equation}\label{28}
E(r)=\dfrac{q}{4\pi\varepsilon_{0}G(\phi)r^{2}}.
\end{equation}
Substituting this solution into equation (\ref{27}) one finds
\begin{align}\label{29}
&-\left[ \dfrac{1}{r^{2}}\dfrac{d}{d r} \left( r^{2}\dfrac{d \phi}{d  r}\right) \right] +\dfrac{T^{2}}{24} \dfrac{\partial V_{\phi\phi}}{\partial\phi}-\dfrac{1}{2}\dfrac{\partial G(\phi)}{\partial\phi} \nonumber\\&\left[\dfrac{q}{4\pi\varepsilon_{0}G(\phi)r^{2}}\right]^{2}+\dfrac{\partial V(\phi)}{\partial\phi}  =0.
\end{align}
Now, considering the fact that $G(\phi)=V(\phi)$ and 
\begin{equation}\label{30}
\lambda=\dfrac{q}{4\pi\varepsilon_{0}},
\end{equation}
we have
\begin{align}\label{31}
&-\left[ \dfrac{1}{r^{2}}\dfrac{d}{d r} \left( r^{2}\dfrac{d \phi}{d  r}\right) \right] +\dfrac{T^{2}}{24} \dfrac{\partial V_{\phi\phi}}{\partial\phi}-\dfrac{\lambda^{2}}{2}\dfrac{\partial V(\phi)}{\partial\phi}\dfrac{1}{V(\phi)^{2}r^{4}}+\nonumber\\& \dfrac{\partial V(\phi)}{\partial\phi}  =0,
\end{align}
which implies
\begin{equation}\label{32}
\left[ \dfrac{1}{r^{2}}\dfrac{d}{d r} \left( r^{2}\dfrac{d \phi}{d  r}\right) \right]=\dfrac{\partial}{\partial\phi}\left[ \dfrac{T^{2}}{24}V_{\phi\phi}+\dfrac{\lambda^{2}}{2}\dfrac{1}{V(\phi)}\dfrac{1}{r^{4}}+V(\phi)\right].
\end{equation}
Now, substituting the potential (\ref{26}) into equation (\ref{32}) gives
\begin{align}\label{33a}
&\left[ \dfrac{1}{r^{2}}\dfrac{d}{d r} \left( r^{2}\dfrac{d \phi}{d  r}\right) \right]=\dfrac{\partial}{\partial\phi}\left[\dfrac{T^{2}}{24}2\left(-\alpha^{2}+3\alpha^{4}\phi^{2}\right) \right]+&\nonumber\\&\lambda^{2}\dfrac{\partial}{\partial\phi}\left[(\alpha^{2}\phi^{2}-1)^{-2}\right] \dfrac{1}{r^{4}}+\dfrac{\partial}{\partial\phi}\left[ \frac{1}{2}(\alpha^{2}\phi^{2}-1)^{2}\right] .
\end{align}

Now, disregarding the term with $\lambda^2$ because we are considering a relatively long distances (far from the charge source $1/r^4$- term), Eq.~(\ref{32}) gives
\begin{equation}\label{33}
\nabla^2\phi=\dfrac{\partial V_{eff}(\phi)}{\partial\phi}.
\end{equation}
Since $V_{eff}(\phi)=V(\phi)+\dfrac{T^2}{24}V_{\phi\phi}$ and $V(\phi)=1/2[(\alpha\phi)^2-1]^2$, 
it follows that
\begin{equation}\label{34}
V_{eff}(\phi)=\dfrac{1}{2}\left[(\alpha\phi)^2-a^2 \right]^2 ,
\end{equation}
where $a^2=1-\dfrac{T^2}{T_{c}^{2}}$ and ${T^{2}_{c}}=\dfrac{4}{\alpha^{2}}.$ The effective potential $V_{eff}$ indicates stability around the new vacuum $\phi_0 =a/\alpha$ for $T<T_c$ (true vacuum) and unstable for $T\geq T_c$ (false vacuum). 

Now perturbing the tachyon fields around its true vacuum $\phi_0$, that is $\phi(r)\rightarrow \phi_0+\eta(r)$, where $\eta(r)$ is the small fluctuation, we can expand (\ref{33}) as
\begin{align}\label{35}
\nabla^2(\phi_0+\eta)&=\dfrac{\partial V_{eff}(\phi)}{\partial\phi}\nonumber\\
&=\nabla^2\phi_0+\nabla^2\eta=\dfrac{\partial V_{eff}}{\partial\phi}|_{\phi_0}+\dfrac{\partial^2 V_{eff}}{\partial\phi^2}|_{\phi_0}\eta \nonumber\\
&\rightarrow\nabla^2\eta=\dfrac{\partial^2 V_{eff}}{\partial\phi^2}|_{\phi_0}\eta.
\end{align}
We have disregarded the terms of higher derivatives because the second derivative is sufficient for our analysis, thus at $\phi_0=a/\alpha$ 
\begin{align}\label{36}
\nabla^2\eta&=4\alpha^2a^2\eta\nonumber\\
&=4\alpha^2\left[1-\dfrac{T^2}{T^2_c} \right]\eta\nonumber\\
&=-4\alpha^2\left[\dfrac{T^2}{T^2_c} -1\right]\eta\nonumber\\
&=-2A\eta , 
\end{align}
where $A=2\alpha^2\left[\dfrac{T^2}{T^2_c}-1 \right]=-2a^2\alpha^2 $. 
Now, developing the Laplacian in Eq.~(\ref{36}) yields 
\begin{equation}\label{37}
\eta''+\dfrac{2}{r}\eta'+2A\eta=0.
\end{equation}
This equation has a solution given by
\begin{equation}\label{39}
\eta(r)=\dfrac{\cosh(\sqrt{2\mid{A}\mid}r)}{\alpha\sqrt{\mid{A}\mid}r},
\end{equation}
where $|A|=-A=2\alpha^2\left(1-\dfrac{T^2}{T^2_c} \right)$. Hence, the dielectric function for this solution is  given as 
\begin{align}\label{40}
G(\phi)&=V(\phi_0+\eta)=V(\phi)|_{\phi_0}+V'(\phi)|_{\phi_0}\eta+\dfrac{1}{2}V''(\phi)|_{\phi_0}\eta^2+{\cal O}(\eta^3)\nonumber\\
&=\dfrac{1}{2}V''(\phi)|_{\phi_0}\eta^2,
\end{align}
where in the last step we went up to second order.
This yields 
\begin{align}\label{41}
G(r)&=2\alpha^2\eta^2\nonumber\\
 &=\dfrac{2}{|A|r^2}\cosh^2\sqrt{2|A|}r.
\end{align}
Substituting this result into the electric field equation modified by dielectric function $G(r)$ we have
\begin{equation}\label{42}
E=\dfrac{\lambda}{r^{2}G(r)}=\dfrac{\lambda}{r^{2}\left[\dfrac{2}{|A|r^2}\cosh^2\sqrt{2|A|}r\right]}. 
\end{equation}
Using the well known relation for determining electric field potential,
$V(r)=\int{E}dr$, to determine the confinement potential $V_{c}(r)$, we get 
\begin{equation}\label{43}
V_{c}(r,T)=\dfrac{\lambda\sqrt{|A|}\tanh(\sqrt{2|A|}r)}{2\sqrt{2}}+c.
\end{equation}
Now, we can compare our equation (\ref{33a}) with the results of \cite{7,23}, for confinement of quarks and gluons 
with $N_{c}$ colors
\begin{align}\label{44}
\dfrac{d^{2}\phi(r)}{d r^{2}}+\dfrac{2}{r}\dfrac{d\phi(r)}{d r}=-\dfrac{g^2}{64\pi^{2}f_{\phi}}\left( 1-\dfrac{1}{N_{c}}\right) \exp\left(- \dfrac{\phi(r)}{f_{\phi}}\right) \dfrac{1}{r^{4}},
\end{align}
thus, equation (\ref{33a}) can be rewritten as
\begin{align}\label{33b}
&\dfrac{d^{2}\phi(r)}{d r^{2}}+\dfrac{2}{r}\dfrac{d\phi(r)}{d r}=\dfrac{\partial}{\partial\phi}\left[\dfrac{T^{2}}{24}2\left(-\alpha^{2}+3\alpha^{4}\phi^{2}\right) \right]+&\nonumber\\&-4\alpha^2\lambda^{2}\left[\phi(\alpha^{2}\phi^{2}-1)^{-3}\right] \dfrac{1}{r^{4}}+\dfrac{\partial}{\partial\phi}\left[ \frac{1}{2}(\alpha^{2}\phi^{2}-1)^{2}\right] .
\end{align}
Since the exponential and quadratic potentials in the former and latter cases are just dielectric functions that modifies the charges, we can now identify our electric charge $q$ in terms of the gluon charge $g$ by comparing the charge source $1/r^4$-terms of both equations (\ref{44}) and (\ref{33b}) to obtain 
\begin{equation}\label{46}
4\lambda^{2}\alpha^2=\dfrac{g^{2}}{32\pi^{2}f_{\phi}}\left( 1-\dfrac{1}{N_{c}}\right).
\end{equation}
Therefore, identifying $\alpha^2=\dfrac{1}{f_{\phi}}$ we find
\begin{equation}\label{47}
\lambda=\dfrac{g}{4\pi}\left( 1-\dfrac{1}{N_{c}}\right)^{\dfrac{1}{2}}, 
\end{equation} 
where we have redefined $g\rightarrow g/2\sqrt{2}$. Using equation (\ref{30}) one can readily find the following relationship between the charges
\begin{equation}\label{48}
q=\varepsilon_{0} g\sqrt{\left( 1-\dfrac{1}{N_{c}}\right)}.
\end{equation}
Substituting the results obtained above into equation (\ref{43}), we have
\begin{align}\label{49}
V_{c}(r,T) &
= \dfrac{g}{4\pi}\sqrt{\left( 1-\dfrac{1}{N_{c}}\right)}\dfrac{\sqrt{|A|} \tanh(\sqrt{2|A|}r)}{2\sqrt{2}}+c.
\end{align}
This represents the static potential observed for the  confinement of quarks and gluons in the tachyon matter. 
At $T=0$ we observe strong confinement regime at short distances. For sufficiently large distances $r$ we observe a steady de-confinement of the quarks and the gluons leading to hadronization. At $T\geq T_c$ the confinement vanishes leading to the breaking of the QCD string tension.

Writing equation (\ref{49}) in a more compact form, we have 
\begin{equation}\label{50}
V_{c}(r,T) = \sigma(T) r  +c. 
\end{equation}  
Where $c$ is the integration constant and $\sigma$ is the QCD string tension which in this case depends explicitly on the temperature. 

The QCD string tension can be written as    
\begin{align}\label{51}
\sigma(T)&\simeq \dfrac{g}{4\pi}\sqrt{\left( 1-\dfrac{1}{N_{c}}\right)}\dfrac{|A|}{2}\nonumber\\
&\simeq \dfrac{g\alpha^2}{4\pi}\sqrt{\left( 1-\dfrac{1}{N_{c}}\right)}\left[1-\dfrac{T^2}{T^2_c} \right] .
\end{align}
 At $T=0$, $\sigma$ does not longer depend on temperature, indicating a constant string tension that binds the quarks together. At this temperature, the quarks and the gluons are automatically in a confined state. At $T=T_{c}$, $\sigma(T=T_{c})$ breaks leading to hadronization. 

Plotting the results from equations (\ref{50}) and (\ref{51}) in Figs.~\ref{Fig1} and \ref{Fig2} we assumed that $\alpha=1$, $\lambda=1$, with this, we get, $g/4\pi=1$, $N_{c}\gg1$ and $c=0$. 

The static potential for the confinement regimes is depicted in Fig.~\ref{Fig1}. At non zero temperatures $T\leq T_{c}$, the potential rises linearly as expected, but the slope decreases with steady increase in temperature from $T=0$ to $T\simeq T_{c}$, where the slope approaches zero. This represents an increase {in the energy} and a decrease in the interactions between the quarks and the gluons as temperature increases. At $T\geq T_{c}$ the confinement vanishes and showing a rise in the energy and a reduction in the interactions of the quarks and the gluons, making them free (asymptotically) in the hadrons. 
Fig.~\ref{Fig2} shows a sharp decrease in $\sigma(T)$ (the coefficient of the linearly increasing potential) with $\sigma $ vanishing at $T=T_{c}$. 

The color dielectric function in Eq.~(\ref{41}) is plotted in Fig.~\ref{Fig3}. As we have earlier shown that $V(r)=G(r)$ we can as well say that $V(r,T)=G(r,T)$. In this sense, we can clearly see from figure ~\ref{Fig1} and ~\ref{Fig3} that the confinement regime/tachyon condensation (at $r\rightarrow r_*$ where $V(r_*)= G(r_*)\rightarrow0$) coincides. We observe from Fig.~\ref{Fig3} that the tachyon condensation corresponds to the minima of the curves $T=0, T_1 ,T_2, T_3, T_4$. It is worth noting that the smaller the minima the higher the depth of the curve and the higher the tachyon condensation, hence we have more tachyons condensing at $T=0$ and as the temperature increases the tachyons becomes gradually free until $T\geq T_c$.
 This regime also coincides with de-confinement phase as seen in Fig.~\ref{Fig1}. 
Thus, by comparing figures  (\ref{Fig1}) and (\ref{Fig3}) we can identify  that 
the electric confinement is associated with tachyon condensation \cite{5}.

\begin{figure}[h!]
  \centerline{\includegraphics[scale=0.7]{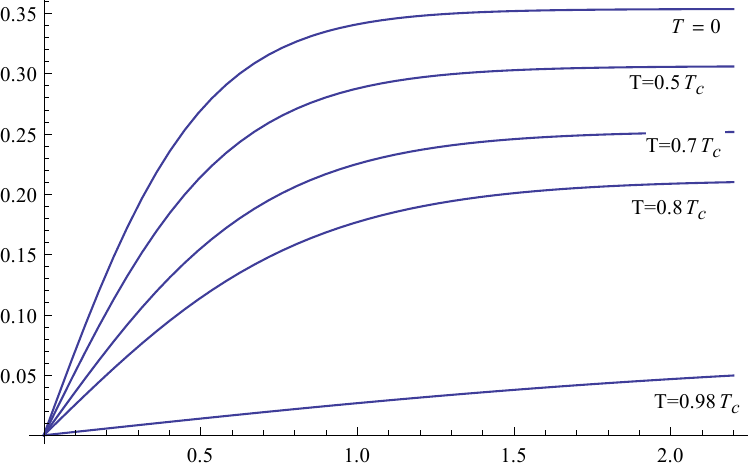}}
  \caption{A plot of a potential $V_{c}(r,T)$ against $r$ for different temperatures.}
  \label{Fig1}
\end{figure}
 \begin{figure}[ht!]
  \centerline{\includegraphics[scale=0.7]{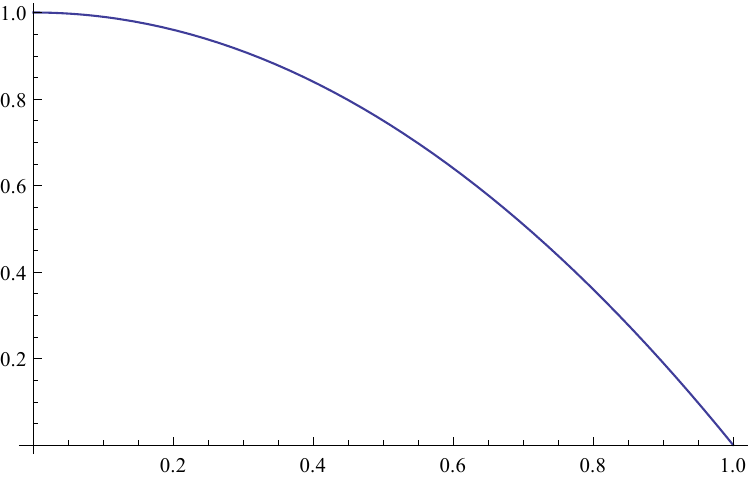}}
  \caption{A plot of the string tension $\sigma(T)$ against $T/T_{c}$.}
  \label{Fig2}
\end{figure}
\begin{figure}[ht!] 
  \centerline{\includegraphics[scale=0.7]{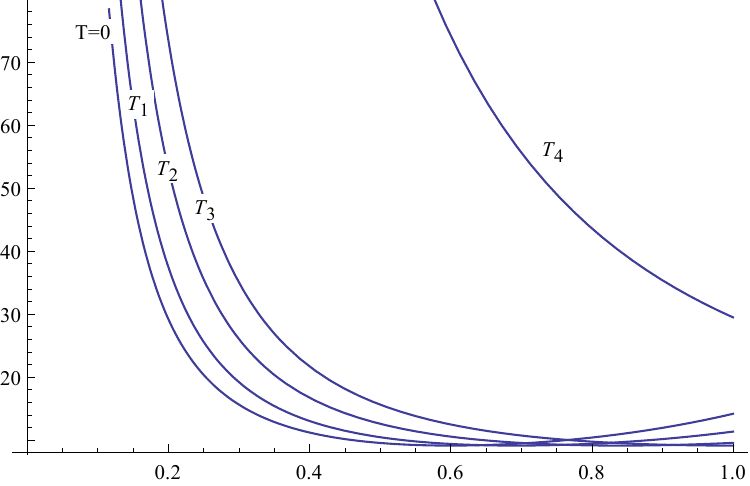}}
  \caption{A plot of the color dielectric function $G(r,T)$ against $r$ for $T_1=0.5T_{c}, T_{2}=0.7T_{c}$, $T_{3}=0.8T_{c}, T_{4}=0.98T_{c}$ and $T=0$.}
  \label{Fig3}
\end{figure}

{\subsection{Glueball Masses}
The search for glueballs has been on for awhile now, unfortunately, the only evidence of its existence is a ``possible" candidate because it has not been confirmed experimentally. They are known to be bound states of pure gluons, mixture of quark and gluon states (hybride), multi-quark bound states etc. Their presence is the consequence of gluon self-interactions in QCD theory. In this section we will be focusing on scalar glueballs. 
They are known to be the lightest in glueball mass, they have QCD degrees of freedom with isospin quantum state of $J^{PC}=0^{++}$, where $J$ is total spin, $P$ is parity and $C$ is charge conjugation with isospin $I=0$ for isoscalars \cite{scalar-glueballs}. 

\subsubsection{ Glueball Mass at $T=0$}\label{gluons at T=0}

From our model, the glueball masses are expected to appear as excitations around the vacuum and are given by \cite{rosina} 
\begin{equation}\label{g1}
M^2_G=\left.\frac{\partial^2 V(\phi)}{\partial\phi^2}\right |_{\phi_0}=4\alpha^2, 
\end{equation}
with $f_\alpha =1/\alpha$ representing the decay constant of the tachyons, hence the glueball mass ($M^2_G$) depends on how fast or slow the tachyon decays. 

\subsubsection{Glueball Mass at a Finite  ($T$)}
We start with Eq.~(\ref{34}) which defines the effective potential of the model, we expand the potential for $\phi\rightarrow\phi_0+\eta $ about the true vacuum of the effective potential, $\phi_0= a/\alpha $, as follows
\begin{align}\label{g2}
V_{eff}(\phi)&=V(\phi)|_{\phi_0}+V'(\phi)|_{\phi_0}\eta+ \frac12V''(\phi)|_{\phi_0}\eta^2+{\cal O}(\eta^3)\nonumber\\
&=\dfrac{1}{2}V''(\phi)|_{\phi_0}\eta^2=2\alpha^2a^2\eta^2.
\end{align}
Comparing the second derivative above with the definition of glueball mass in Eq.~(\ref{g1}) we arrive at
\begin{align}\label{g3}
M^2_G(T)&=4\alpha^2 a^2\nonumber\\
&=M^2_G(0) \left[1-\frac{T^2}{T^2_c} \right].
\end{align}
Hence, at $T=0$ we retrieve Eq.~(\ref{g1}) and at $T=T_c$, $M^2_G(T)$ vanishes. This model seems to show a remarkable resemblance with the lattice simulation results. The glueballs should be seen as ``rings of glue" which are kept together by the string tension $\sigma(T)$ contained in the interquark potential and vanishes at $T=T_c$ when the string breaks depicting de-confinement phase. At $T<T_c$ the thermodynamic properties of the model can be well understood and studied in terms of gas of glueballs with $M^2_G(T)< M^2_G(0)$ \cite{glueballs-T}.

We may now establish a relationship between string tension and glueball masses through equations (\ref{g1}) and (\ref{51}) at $T=0$, i.e.,
\begin{eqnarray}\label{51-t}
\sigma(0)
\simeq \dfrac{gM^2_G(0)}{16\pi}\sqrt{ 1-\dfrac{1}{N_{c}}}.
\end{eqnarray}
Recalling that $g=\sqrt{16\pi\alpha_s/3}$ is the chromoelectric charge, where $\alpha_s=0.45$ is close to the QCD coupling constant, and assuming $N_c=3$, we find the expected results $\sqrt{\sigma{(0)}}=420$ MeV for a glueball mass $M_G(0)=1184$ MeV \cite{rosina}.
}

To end this section, few comments in connection with  Sec.~\ref{rgd} are in order. 
Substituting the effective potential (\ref{g2}) into the gluon condensate in Eq.~(\ref{gd7}) we get,
\begin{align}\label{g5}
\langle G'F^{\mu\nu}F_{\mu\nu}\rangle&=4|\epsilon_v|\langle 2\alpha^2a^2\eta^2-1 \rangle\nonumber\\&
=4|\epsilon_v|\langle 2\alpha^2\left[1-\dfrac{T^2}{T^2_c} \right] \eta^2-1 \rangle.
\end{align}
Hence, the gluon condensate decreases with increasing temperature and vice versa. At $T=T_c$ we recover the well-known gluon condensate at zero temperature
 \begin{equation}\label{g6}
\langle G'F^{\mu\nu}F_{\mu\nu}\rangle=-4|\epsilon_v|.
\end{equation} 

\section{Conclusions}
\label{sec4}

In our investigations we find the net static potential for confinement phase of quarks and gluons as a function of temperature. We used the Abelian QED theory to approximate the non-Abelian QCD theory. We do this  by employing a phenomenological effective field theory involving tachyon field dynamics coupled to electromagnetism via color dielectric function.
The color dielectric function is responsible for the long distance interactions to bring about confinement in the infrared (IR) regime. It also modifies the gluon condensate $\langle G'F^{\mu\nu}F_{\mu\nu}\rangle $, develops tachyon condensation and consequently allows confinement in the IR regime. We show that the confinement is favored at short distances and low temperatures, whereas de-confinement shows up at long distances and higher temperatures in the tachyon matter.
Confinement of quarks and gluons coincides with tachyon condensation within the same temperature ranges as it is shown in Fig.~\ref{Fig1} and \ref{Fig3}. As a result, de-confining phase at $T\geq T_{c}$ does not correspond to the tachyon condensation as it is seen in Figs.~\ref{Fig1} and \ref{Fig3}. Consequently, tachyon condensation is associated with electric field confinement and thus, our results conform with QCD monopole condensation as predicted in the well-known dual scenario \cite{dual}.  
In such dual scenario, the QCD-monopole condensation is necessary for spontaneous chiral-symmetry breaking \cite{dual,27,28,29}. 
Thus, in our setup it is expected that at the confining phase ($T<T_c$) there is a spontaneous chiral-symmetry breaking, whilst at the de-confining phase ($T\geq T_{c}$) there is a restoration of the chiral-symmetry.  

The QCD string tension and scalar glueball mass were also computed as a function of temperature. They both decrease rapidly with temperature and break (vanish) at $T=T_{c}$.
Finally, we intend to advance further studies in this subject by studding confinement of fermionic tachyons using similar approach.
 
\acknowledgments

We would like to thank CNPq and CAPES for partial financial support.

\end{document}